\documentclass{eptcs}
\providecommand{\event}{TFPiE 2025} 

\usepackage{iftex}

\ifpdf
  \usepackage{underscore}         
  \usepackage[T1]{fontenc}        
\else
  \usepackage{breakurl}           
\fi

\usepackage{mathtools}
\usepackage{amssymb}
\usepackage{stmaryrd}

\usepackage{tabularray}
\usepackage{makecell}
\usepackage{leftindex}

\usepackage{float}
\floatstyle{boxed}
\restylefloat{figure}

\usepackage{tikz}
\usetikzlibrary{scopes,positioning,calc,fit,arrows,automata}

\usepackage{tikzit} 

\tikzstyle{Normal Box}=[shape=rectangle, rounded corners, draw=black, very thick, inner sep=5pt, minimum height=1cm, text centered, fill=white, minimum width=0.75cm, tikzit category=Operations, align=center]
\tikzstyle{Normal Meta Box}=[shape=rectangle, rounded corners, draw=black, dashed, very thick, inner sep=5pt, minimum height=1cm, text centered, fill=white, minimum width=0.75cm, tikzit category=Operations]
\tikzstyle{Tall Box}=[shape=rectangle, rounded corners, draw=black, very thick, inner sep=5pt, minimum height=2cm, text centered, fill=white, minimum width=0.75cm, tikzit fill={rgb,255: red,213; green,255; blue,27}, tikzit category=Operations, align=center]
\tikzstyle{Tall Meta Box}=[shape=rectangle, rounded corners, draw=black, dashed, very thick, inner sep=5pt, minimum height=2cm, text centered, fill=white, minimum width=0.75cm, tikzit fill={rgb,255: red,213; green,255; blue,27}, tikzit category=Operations]
\tikzstyle{Very Tall Box}=[shape=rectangle, rounded corners, draw=black, very thick, inner sep=5pt, minimum height=3cm, text centered, fill=white, minimum width=0.75cm, tikzit fill={rgb,255: red,29; green,173; blue,3}, tikzit category=Operations]
\tikzstyle{spacing node}=[fill=none, draw=none, shape=circle, tikzit fill={rgb,255: red,191; green,128; blue,64}, minimum width=5mm, minimum height=5mm, tikzit category=Operations]
\tikzstyle{Input Dot}=[fill=black, draw=black, shape=circle, tikzit category=Idioms, inner sep=2pt, outer sep=2pt]
\tikzstyle{Silent Input Dot}=[draw=black, shape=circle, tikzit category=Idioms, inner sep=2pt, outer sep=2pt]
\tikzstyle{Output}=[fill=none, draw=black, shape=rectangle, tikzit category=Idioms, outer sep=2pt, inner sep=2pt, anchor=mid west, minimum height=1.5em]
\tikzstyle{Silent Output}=[fill=none, draw=black, shape=rectangle, tikzit category=Idioms, outer sep=2pt, inner sep=2pt, anchor=mid west, minimum height=1.5em, dashed, tikzit fill=blue]
\tikzstyle{Input Port}=[tikzit category=Idioms, anchor=south, tikzit fill={rgb,255: red,17; green,255; blue,0}]
\tikzstyle{Output Port}=[tikzit category=Idioms, anchor=north, tikzit fill={rgb,255: red,255; green,0; blue,4}]
\tikzstyle{Text node}=[tikzit category=Idioms, tikzit fill={rgb,255: red,242; green,0; blue,255}, anchor=mid west, align=center]
\tikzstyle{multiline}=[align=center]

\tikzstyle{Edge}=[>=stealth, ->]
\tikzstyle{Dashed Edge}=[-, dashed]
\tikzstyle{round}=[-, rounded corners, tikzit draw=blue]
\tikzstyle{order edge}=[draw=red, ->, thick]
\tikzstyle{order edge dashed}=[draw=red, ->, thick, dashed]
\tikzstyle{brace-line}=[-, decorate, tikzit draw={rgb,255: red,255; green,128; blue,0}]
\tikzstyle{Dashed Arrow}=[dashed, ->, >=stealth]

\input{diagrams/diagrams.tikzdefs}

\tikzset{
  outside/.style={
    rectangle,
    rounded corners,
    draw=black,
    very thick,
    minimum height=2em,
    inner sep=6pt,
    text centered,
  },
  outside-no-xsep/.style={
    outside,
    inner xsep=-3pt,
  },
  inputDot/.style={
    fill,
    circle,
    inner sep=2pt,
    outer sep=2pt,
  },
}

\newsavebox{\genericfilt}
\newcommand{\insertDiagramWithBorderScaleAnchor}[5]{
  \savebox{\genericfilt}{\input{#4.tikz}}
  \begin{tikzpicture}[baseline=(current bounding box.#2)]
    \node [#3,label=below:{#5},scale=#1]  {\usebox{\genericfilt}};
  \end{tikzpicture}
  }

\newcommand{\insertDiagramWithBorderScale}[4]{\insertDiagramWithBorderScaleAnchor{#1}{mid west}{#2}{#3}{#4}}
\newcommand{\insertDiagramWithBorderAnchor}[4]{\insertDiagramWithBorderScaleAnchor{0.75}{#1}{#2}{#3}{#4}}
\newcommand{\insertDiagramWithBorder}[3]{\insertDiagramWithBorderAnchor{mid west}{#1}{#2}{#3}}

\newcommand{\metaName}{abstract pattern}
\newcommand{\metaNameCap}{Abstract pattern}

\newcommand{\oliver}[1]{\textcolor{red}{[Oliver: #1]}}
\renewcommand{\oliver}[1]{}

\title{Intent Preserving Generation of Diverse and Idiomatic (Code-)Artifacts}
\author{Oliver Westphal
\institute{Universität Duisburg-Essen}
\email{oliver.westphal@uni-due.de}
}

\def\titlerunning{Intent Preserving Generation of Diverse and Idiomatic (Code-)Artifacts}
\def\authorrunning{Oliver Westphal}

\begin{document}
  \maketitle

\begin{abstract}
  When automatically generating programming exercise tasks one often also needs to automatically generate programs.
  At the very least when providing sample solutions is part of automated feedback.
  But programs can also be used as part of the exercise task description to communicate a task's requirements~\cite{wflp20-paper}.

  Writing good program generators that produce varied yet idiomatic code while being easily adaptable for new tasks is challenging.
  The challenges are intensified if task generation requires additional artifacts, like a more general behavior specification for testing or additional textual descriptions.
  Manually writing generators for multiple different but strongly related artifacts gets complicated quickly.

  We present an approach where instead of writing monolithic generators for multiple connected artifacts one specifies a small set of abstract building blocks and for each such building block defines sets of concrete realizations for various kinds of artifacts.
  Then the intended structure of the resulting artifacts is specified as a composition of the small abstract building blocks.
  This abstract description then serves as the common source from which related artifacts can be derived automatically.
  The approach is generic in the kind of artifacts it can produce and is therefore adaptable to a wide range of contexts.
\end{abstract}

\section{Introduction}
\label{sec:intro}

Automatic generation of exercise tasks is beneficial in many ways.
It reduces the workload of educators and allows for on-demand training sessions for students.
Especially when combined with an automatic grading system.
We have previously presented a specification language for console IO-behavior that enables property-based testing of console IO-programs~\cite{tfpie19-paper} improving automatic grading capabilities for this class of programs.
Additionally, we have developed a framework for building parameterized templates for exercise tasks~\cite{wflp20-paper}.
An exercise template is essentially a function from some parameter type $p$ to a task description and test suite.
Together with a (random) generator for values of type $p$ these templates allow for quick and automatic creation of new exercise tasks.
Within this framework we have explored the use of program code to communicate task requirements and to produce variability in automatically generated exercise tasks on Haskell IO-programming.
Expressing requirements through code has the advantage that one does not need to produce precise natural language descriptions when generating randomized variants of exercise tasks.

For example, we can give students a randomly generated program and ask for the observable IO-behavior for a given sequence of inputs.
Or we give two syntactically different programs and ask them whether these programs have the same observable IO-behavior.
Both of these tasks focus on testing student's program comprehension skills, but we can also use randomly generated programs to create tasks testing program writing skills.
For example, we can give students a program and require them to write a program with identical IO-behavior, but using a different solution strategy (using a specific higher-order function etc.).
For such tasks we do not only need to generate a random program but also test-cases in form of a specification describing the correct program behavior (using the language of~\cite{tfpie19-paper}).

The quality of randomized instantiations of such task templates depends on the quality of the programs we can generate automatically.
Naive random generation of syntactically valid programs is possible, but invariants are difficult to enforce.
Especially since we usually have a general idea for the structure of programs/behavior we want for a specific task.
For example, we might want programs that interactively read in a sequence of values into a list, compute something on that list and finally report the result of that computation.
Different variations of such (simple) program blueprints then lead to variants of tasks with different levels of difficulty or learning goals.
Additionally, naive randomized generation of programs rarely yields programs that resemble human-written code.
In other words they are not idiomatic.
It is however important to teach using examples highlighting best practices early on.

Additional complexity arises when we need not just a single program but multiple related artifacts, e.g., a specification of IO-behavior, code in another language or a supporting verbal description.
Essentially, the generation procedure now needs to be aware of the intended meaning of a generated artifact so synchronization across multiple artifacts becomes possible.
Generation of behavior (specifications) and programs in tandem is tedious and so is (partially) translating one into the other.
Instead, we aim to derive random programs, behavior, and any other required artifact from a common and more abstract source.

We present an approach where different artifacts can be generated from a single description of desired data-flow structure.
We will mostly focus on generating Haskell IO programs, but the approach is not restricted to specific artifact kinds.
Our goal is to have a framework in which we can specify how basic abstract operations can be translated to various kinds of partial artifacts.
Possibly with multiple different versions per targeted artifact.
From a composition of basic abstract operation we then generate the different kinds of possible artifacts automatically.
With this approach ensuring semantic coherence of different artifacts still is the responsibility of the educator, but deriving from a shared description and the compositional structure of this common source makes artifacts coherence more manageable (see Section~\ref{sec:coherence} for more details).
The underlying combinatorics of a modular approach also allow us to easily create larger numbers of variations in generated artifacts compared to writing monolithic generators.

We intend to use the presented system, in combination with the exercise task templates from our previous work, to enrich the Haskell IO-tasks we use in our second year course teaching programming paradigms.
The introduction of monadic IO is a common stumbling block for novice Haskell programmers, especially those with previous experience in imperative languages.
Being able to quickly create and vary exercise tasks will allow us to provide students with exactly the right amount of practice material they require, without ballooning the time we spend designing tasks.

\section{Overview}

In order to generate multiple related artifacts automatically, we first need to decide how to specify the shared intent of these artifacts.
By intent, we mean the general high-level idea for the structure of some computation.
Additionally, a single intent is meant to describe a single specific behavior and not a class of related behaviors.
Let us take as example the intent of first reading a natural number $n$ from the console, followed by reading in $n$ additional numbers, then summing up these $n$ values, and finally reporting back the summation's result.

How can we describe intent like this in a way that is useful for generating different programs implementing this behavior as well as other related artifacts?
The example intent is a composition of four smaller intents and ideally our description of intent should also be composed out of smaller building blocks in the same way.

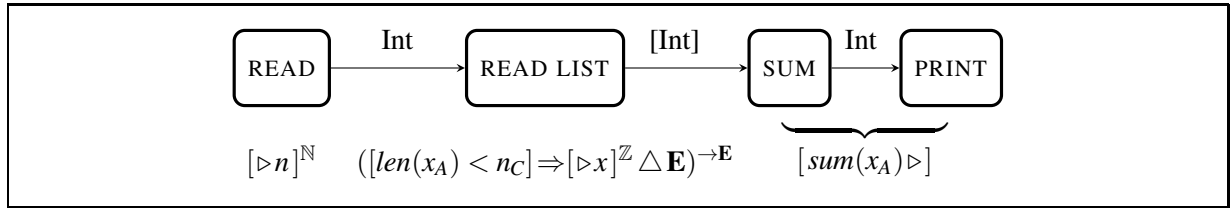
\begin{figure}
  \medskip
  \begin{center}
    \begin{tikzpicture}
	\begin{pgfonlayer}{nodelayer}
		\node [style=Normal Box] (0) at (-7.5, 0) {\idiomName{read}};
		\node [style=Normal Box] (1) at (-0.5, 0) {\idiomName{read list}};
		\node [style=Normal Box] (2) at (6, 0) {\idiomName{sum}};
		\node [style=Normal Box] (3) at (10.3, 0) {\idiomName{print}};
		\node [style=none] (4) at (-4.425, 0.75) {Int};
		\node [style=none] (5) at (7.9, 0.75) {Int};
		\node [style=none] (6) at (2.95, 0.7) {[Int]};
		\node [style=none] (7) at (-7.5, -2.5) {\readInput{n}{\nat}};
		\node [style=none] (8) at (-0.5, -2.5) {$(\branch{\mathit{len}(x_A) < n_C}{\readInput{x}{\integer}}{\loopExit})^\loopArr$};
		\node [style=none] (9) at (8, -2.5) {\writeOutputSimple{\mathit{sum}(x_A)}};
		\node [style=none] (10) at (8, -1.25) {$\underbrace{\phantom{\writeOutput{\mathit{sum}(x_A)}}}$};
	\end{pgfonlayer}
	\begin{pgfonlayer}{edgelayer}
		\draw [style=Edge] (0) to (1);
		\draw [style=Edge] (1) to (2);
		\draw [style=Edge] (2) to (3);
	\end{pgfonlayer}
\end{tikzpicture}
  \end{center}
  \caption{\label{fig:example-with-spec} Data-flow diagram and corresponding specification expression}
\end{figure}

We could describe intent with a specification expression in the language from our previous work~\cite{tfpie19-paper}.
Figure~\ref{fig:example-with-spec} shows such an expression for the example intent alongside a wiring diagram representing the data-flow of the verbal description.
The boxes in the data-flow represent the conceptual steps present in the verbal description and wires depict the data required and provided by each operation.
For example, reading the list of $n$ integers (\idiomName{read list}) produces an integer list and requires another operation to provide the value for $n$.
Specifications are structured similarly to regular expressions.
Input actions like $\readInput{n}{\nat}$ describe reading in a new value from the given set into a variable (here reading a natural number into $n$).
Variables store the sequence of values read into them in chronological order.
Output actions like $\writeOutputSimple{\mathit{sum(x_A)}}$ specify that the result of evaluating some term is written to the console.
Here, the output is the sum of all values read into $x$.
Instead of accessing \textbf{a}ll values with the $A$-subscript, terms can also access the most recent value of some variable with the $C$-subscript (for \textbf{c}urrent).
The specification language has two control structures. $\branch{\cdot}{\cdot}{\cdot}$ for branching, where the condition in brackets gives control to the expression left of $\triangle$ if it evaluates to true or to the right expression otherwise.
$(\dots)^\loopArr$ iterates the enclosed specification until inside it the exit marker $\loopExit$ is encountered.

Specification subexpressions in Figure~\ref{fig:example-with-spec} are aligned with boxes in the data-flow showing which parts of the specification encode which part of the overall behavior.
The specification expression completely describes the IO-behavior we are interested in, so at first glance it seems to be a good starting point for our generation process.
We could then try to translate such an expression into one or multiple different programs.
However, straightforward translation into a program results in an ad-hoc re-implementation of the specification's semantics (see~\cite{flops20-paper} for such a translation).
Such programs are usually highly non-idiomatic and therefore not what we want.

For higher quality programs, we need to know more about the intent than just its IO-behavior.
In essence, generating idiomatic programs for a given specification requires us to infer composition boundaries and the intended meaning of subexpressions.
We need information on how the different parts that make up the whole specification interface with each other.
That is, we need to infer something similar to the data-flow diagram in Figure~\ref{fig:example-with-spec} from a given specification expression.

In our experience such an analysis gets complicated very quickly and heavily relies on anticipating and recognizing handcrafted usage patterns that signal intent.
It is therefore more efficient to simply start with the data-flow building blocks directly.
In fact, they are usually already present (implicitly) in the instructor's mind anyway.
We therefore describe intent as data-flow wiring diagrams that we will introduce in detail in the Section~\ref{sec:abstract}.
We will refer to both the boxes inside the diagram and the diagram as a whole as intent, since the individual boxes can be replaced by other diagrams resulting in a more detailed description of the desired behavior.

Explicitly stating intent, additionally, gives instructors fine-grained control over the result of the artifact generation.
For example, we can easily change the \idiomName{sum} to a \idiomName{product} simply by replacing the box in the diagram.
Given just a specification expression such replacements would again require careful decomposition.

Now, how can we get from a data-flow diagram to idiomatic programs, or any other kind of artifact?
For each operation, i.e., the boxes in the diagram, we can find one or more idiomatic implementations.
For example, reading in a list of $n$ values from the console can be implemented in different ways in Haskell.
One can, for example, use a higher order function like \lstinline|replicateM|.
\begin{lstlisting}
  xs <- replicateM n readLn
\end{lstlisting}
Another common pattern is to use explicitly recursive functions.\\
  \begin{minipage}{.55\textwidth}
    \begin{lstlisting}
  let
    go list
      | length list == n = pure list
      | otherwise = do
          x <- readLn
          go (list++[x])
  xs <- go []
    \end{lstlisting}
  \end{minipage}
  \begin{minipage}{.4\textwidth}
    \begin{lstlisting}
  let
    go 0 list = pure list
    go i list = do
      x <- readLn
      go (i-1) (list++[x])
  xs <- go n []
    \end{lstlisting}
  \end{minipage}

All the above fragments require a value $n$ to be in scope and in turn introduce a new identifier $xs$ that holds the read list.
This corresponds exactly to the data-flow into and out of the \idiomName{read list} box.
Given program fragments for the other boxes in the same way, we can then combine them into a valid complete program for the described intent.
By providing multiple concrete realizations per abstract box the resulting programs can cover different possible idiomatic programs adhering to the original intent.
We will introduce such concrete realizations in more detail in Section~\ref{sec:concrete}.

In addition to fragments for a single intent building block, we can also have fragments that realize two or more basic intents at the same time.
For example, the operation of summing up the values in the list can be combined into the second recursive loop function, eliminating the intermediate list.
Realization of intent can therefore depend on its surrounding context.

\begin{figure}[b]
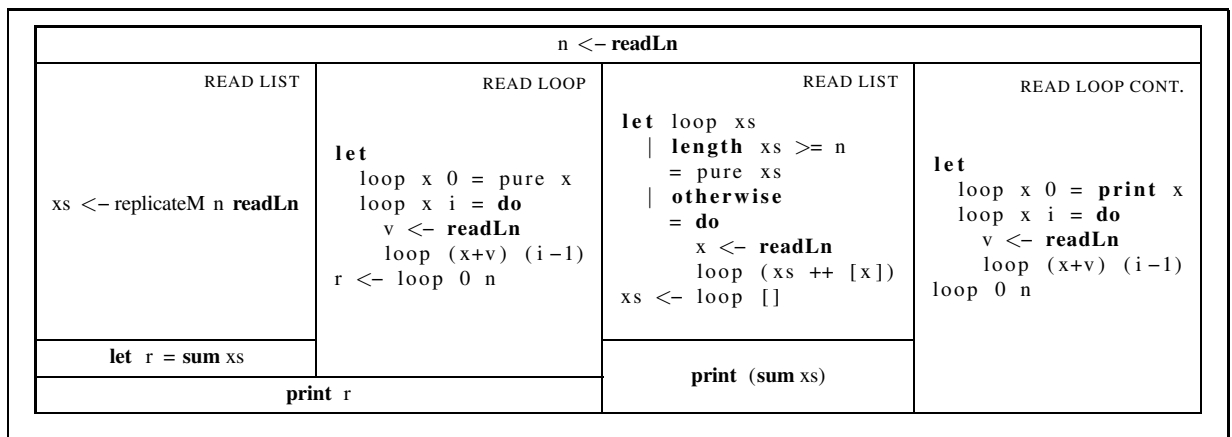

  \smallskip
  \begin{center}
  \scriptsize
    \begin{tblr}{
        vlines,
        hlines,
        cells = {c,t},
        cell{1}{1} = {c=4}{},
        cell{2}{2} = {r=2}{},
        cell{2}{4} = {r=3}{},
        cell{3}{3} = {r=2}{},
        cell{4}{1} = {c=2}{},
      }
      \lstinline|n <- readLn|  &   &   &   \\
        {\hfill \idiomName{read list}\vspace{1.3cm}\\ \lstinline|xs <- replicateM n readLn|}
        & {\hfill \idiomName{read loop}\vspace{4mm}\\ 
\begin{lstlisting}[firstline=10, lastline=15]
let
  loop x 0 = pure x
  loop x i = do
    v <- readLn
    loop (x+v) (i-1)
r <- loop 0 n
\end{lstlisting} \vspace{7mm}}
        & {\hfill \idiomName{read list}\\ 
\begin{lstlisting}[firstline=1, lastline=8]
let loop xs
  | length xs >= n
    = pure xs
  | otherwise
    = do
      x <- readLn
      loop (xs ++ [x])
xs <- loop []
\end{lstlisting}}
        & {\hfill \idiomName{read loop cont.}\vspace{5mm}\\ 
\begin{lstlisting}[firstline=17, lastline=22]
let
  loop x 0 = print x
  loop x i = do
    v <- readLn
    loop (x+v) (i-1)
loop 0 n
\end{lstlisting}\vspace{10mm}} \\
      \lstinline|let r = sum xs| &   & \lstinline|print (sum xs)| &   \\
      \lstinline|print r| &   &   &   \\
    \end{tblr}
  \end{center}
  \caption{\label{fig:idiom-table} Possible idiomatic realizations leading to six different programs}
\end{figure}

Figure~\ref{fig:idiom-table} shows a few possible realizations of our example intent.
Two concrete realizations for the same operation can be used interchangeably.
In this case the two fragments labeled \idiomName{read list} can replace each other allowing us to build a total of six slightly different idiomatic programs from the given fragments.
All of these programs have the same observable IO-behavior.

This procedure of defining realizations for each of the boxes and then combining them according to the specified data-flow works for arbitrary kinds of artifacts.
In the same way we derive programs, we can also derive the specification expression in Figure~\ref{fig:example-with-spec} and in fact even the verbal description of our intent (see \ref{sec:other} for more details).

\section{Describing intent}
\label{sec:abstract}

At the center of our framework are what we call abstract implementations.
An abstract implementation is a representation of intent built from one or more abstract idioms.
Conceptually, abstract idioms represent \emph{exactly} one operation or action.
That means, the realization of an abstract idiom, and by extension of an abstract implementation, is only allowed to choose \emph{how} the intent is realized but not \emph{what} the intent actually is (see Section~\ref{sec:meta} for also choosing the ``what'').

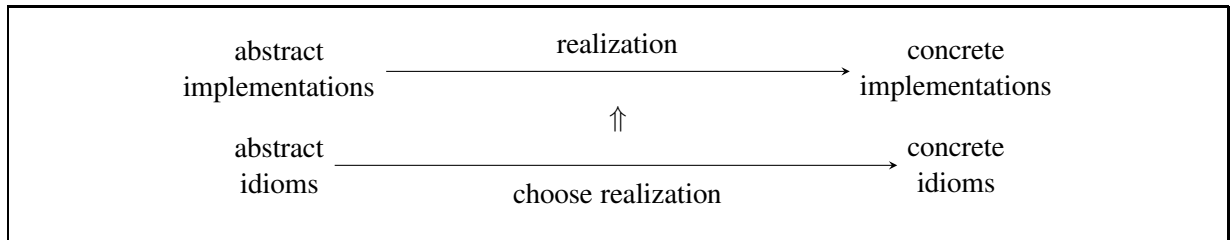
\begin{figure}[t]
  \begin{center}
    \begin{tikzpicture}
	\begin{pgfonlayer}{nodelayer}
		\node [style=multiline] (0) at (-9, 1.25) {abstract\\implementations};
		\node [style=multiline] (1) at (9, 1.25) {concrete\\implementations};
		\node [style=multiline] (2) at (-9, -1.25) {abstract\\idioms};
		\node [style=multiline] (3) at (9, -1.25) {concrete\\idioms};
		\node [style=multiline] (4) at (0, 0) {$\Uparrow$};
		\node [style=multiline] (7) at (0, -2) {choose realization};
		\node [style=none] (8) at (0, 2) {realization};
		\node [style=spacing node] (9) at (0, 2.25) {};
	\end{pgfonlayer}
	\begin{pgfonlayer}{edgelayer}
		\draw [style=Edge] (0) to (1);
		\draw [style=Edge] (2) to (3);
	\end{pgfonlayer}
\end{tikzpicture}
  \end{center}
  \caption{\label{fig:abstract-concrete}Lifting of idiom realization.}
\end{figure}

Abstract implementations can be seen as a rough outline or a design recipe for a concrete artifact, e.g.,  a concrete program.
In order to construct such an artifact we can define a mapping from abstract idioms to concrete ones, e.g., to concrete program fragments, and then use the structure of the abstract implementation, i.e., how different abstract idioms are connected, to obtain a program following the design recipe represented by the abstract implementation (see Figure~\ref{fig:abstract-concrete}).

Abstract idioms have a typed interface of inputs and outputs and abstract implementations are built by composing multiple abstract idioms along those interfaces.
We use abstract implementations as the starting point from which we derive the various concrete artifacts, including programs and specification expressions.

In this section we describe the structure of abstract idioms and how they are composed into abstract implementations.
We define two transformations for systematically manipulating abstract implementations.
With these transformations we show how to introduce intent preserving variations in data-flow, including context-sensitive transformations.

\subsection{Abstract implementations}

Abstract implementations are built from abstract idioms.
Each abstract idiom is given a signature consisting of the types of its inputs and outputs together with a label that represents the intent of the operation.
Additionally, we mark idioms whose realizations as program fragments perform IO-side-effects.
We denote abstract idioms as labeled boxes with arrows for the inputs on the left and outputs on the right.

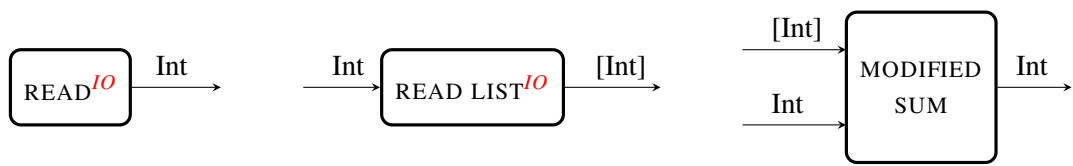
\begin{center}
  \begin{tikzpicture}
	\begin{pgfonlayer}{nodelayer}
		\node [style=Normal Box] (0) at (0, 0) {\idiomName{read\markSideEffect}};
		\node [style=none] (2) at (4, 0) {};
		\node [style=Text node] (4) at (2, 0.5) {Int};
	\end{pgfonlayer}
	\begin{pgfonlayer}{edgelayer}
		\draw [style=Edge] (0) to (2.center);
	\end{pgfonlayer}
\end{tikzpicture}

  \qquad
  \begin{tikzpicture}
	\begin{pgfonlayer}{nodelayer}
		\node [style=Normal Box] (0) at (0, 0) {\idiomName{read list\markSideEffect}};
		\node [style=none] (1) at (-4.5, 0) {};
		\node [style=none] (2) at (5, 0) {};
		\node [style=Text node] (3) at (-4, 0.5) {Int};
		\node [style=Text node] (4) at (3, 0.5) {[Int]};
	\end{pgfonlayer}
	\begin{pgfonlayer}{edgelayer}
		\draw [style=Edge] (1.center) to (0);
		\draw [style=Edge] (0) to (2.center);
	\end{pgfonlayer}
\end{tikzpicture}
  \qquad
  \begin{tikzpicture}
	\begin{pgfonlayer}{nodelayer}
		\node [style=Tall Box] (0) at (0, 0) {\idiomName{modified} \\ \idiomName{sum}};
		\node [style=none] (1) at (-4.75, 1) {};
		\node [style=none] (2) at (4, 0) {};
		\node [style=Text node] (3) at (-4.25, 1.5) {[Int]};
		\node [style=Text node] (4) at (2.25, 0.5) {Int};
		\node [style=none] (5) at (-2, -1) {};
		\node [style=none] (6) at (-2, 1) {};
		\node [style=none] (7) at (-4.75, -1) {};
		\node [style=Text node] (8) at (-4.25, -0.5) {Int};
	\end{pgfonlayer}
	\begin{pgfonlayer}{edgelayer}
		\draw [style=Edge] (0) to (2.center);
		\draw [style=Edge] (1.center) to (6.center);
		\draw [style=Edge] (7.center) to (5.center);
	\end{pgfonlayer}
\end{tikzpicture}
\end{center}

As already stated above, the labels of abstract idioms represent exactly one specific operation.
The intent behind the first and second idiom is relatively clear from the types and name alone.
They are meant to represent reading a single value and a sequence of reads with length equal to the input parameter, respectively.
For the third idiom it is less clear, but it nonetheless must represent exactly one operation and not a class of different modified sums.
We here define it to mean summing up a list but adding all values equal to the second parameter twice.

Note that the types used in the signatures of abstract idioms are as abstract as the labels, but in most cases we will name them after Haskell types.

Now abstract implementations are built by connecting multiple abstract idioms in a way compatible with their signatures.
An abstract implementation is therefore a wiring diagram connecting abstract idioms.

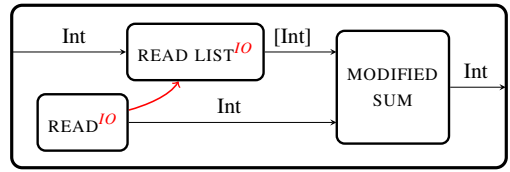
\begin{center}
\insertDiagramWithBorder{outside-no-xsep,inner ysep=8pt}{diagrams/abstract-op/singleComposition}{}
\end{center}

We require this diagram to be acyclic.
In addition to the black wires denoting data-flow between abstract idioms, each diagram defines the exact ``execution'' order of IO-side-effects as red wires.
This ordering guarantees that the abstract implementation describes exactly one behavioral pattern.

The diagram's effect order $\prec_{\mathit{IO}}$ is a total order on the subset of abstract idioms with IO-side-effects that is compatible with the data-flow relation $\prec_{\mathit{data}}$ induced by the black wiring, i.e., the red and black wires together must not form any cycles.
For both relations we define $x \prec y$ to mean that we can go from $x$ to $y$ in the diagram via a (possibly empty) sequence of the arrows for that relation.
Both orders are compatible iff for all abstract idioms $x$ and $y$, if $x \prec_{\mathit{data}} y$ and $x \neq y$ then $y \not\prec_{\mathit{IO}} x$.

We define two basic transformations to manipulate abstract implementations:
A substitution replacing an abstract idiom with a more detailed diagram with the same outside interface, and a transformation that finds an occurrence of a given pattern in an abstract implementation and merges this pattern into a single abstract idiom.

\begin{figure}[t]
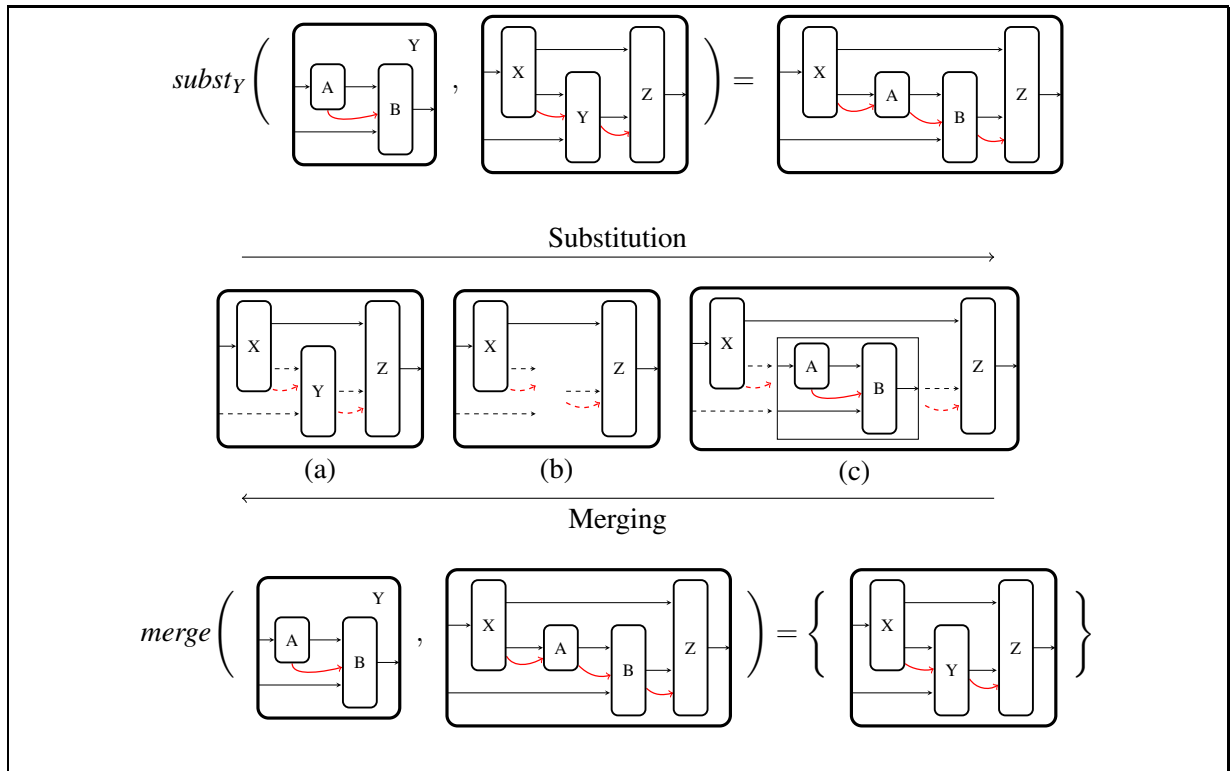

  \caption{\label{fig:steps}Intermediate steps of substitution (top) and merging (bottom) operations}
  \begin{center}
    $\mathit{subst}_Y\Biggl($
    \insertDiagramWithBorderScale{0.6}{outside-no-xsep}{diagrams/abstract-op/small}{}
    $,$
    \insertDiagramWithBorderScale{0.6}{outside-no-xsep}{diagrams/abstract-op/large}{}
    $\Biggr)=$
    \insertDiagramWithBorderScale{0.6}{outside-no-xsep}{diagrams/abstract-op/result}{}
  \end{center}

  \begin{center}
    \tikz\draw[->] (0,0) -- node[above]{Substitution} (10,0);\\[2ex]
    \insertDiagramWithBorderScale{0.6}{outside-no-xsep}{diagrams/abstract-op/large-marked}{(a)}
    \insertDiagramWithBorderScale{0.6}{outside-no-xsep}{diagrams/abstract-op/large-deleted}{(b)}
    \insertDiagramWithBorderScale{0.6}{outside-no-xsep}{diagrams/abstract-op/large-filled}{(c)}
    \tikz\draw[<-] (0,0) -- node[below]{Merging} (10,0);
  \end{center}

  \begin{center}
    $\mathit{merge}\Biggl($
    \insertDiagramWithBorderScale{0.6}{outside-no-xsep}{diagrams/abstract-op/small}{}
    $,$
    \insertDiagramWithBorderScale{0.6}{outside-no-xsep}{diagrams/abstract-op/result}{}
    $\Biggr)=\Biggl\{$
    \insertDiagramWithBorderScale{0.6}{outside-no-xsep}{diagrams/abstract-op/large}{}
    $\Biggr\}$
  \end{center}
\end{figure}

Figure~\ref{fig:steps} gives an example for both transformations and shows their intermediate steps.
For substitution, we first cut all wires into and out of the idiom that is to be replaced (a) and remove it from the diagram (b).
We then insert the replacement diagram (c), connect the dangling wires and remove the outside border of the inserted diagram.
Note that substitution always replaces exactly one abstract idiom.
If multiple idioms with the same label exist in the diagram the subscript of $\mathit{subst}$ must clearly identify the one to be replaced.
Merging of a pattern into a single abstract idiom is done by first finding an occurrence of the pattern in the diagram and then performing the intermediate steps of the substitution process in reverse.
We start by cutting the wires at the boundary of the pattern occurrence (c), then remove the pattern from the diagram (b), insert a new idiom with the outside label of the argument pattern (a), and finally reconnect the dangling wires.
Note that a pattern might have multiple different matches in a diagram.
Merging therefore returns a set of new diagrams.

In order for merging to produce a well-formed abstract implementation the match in the surrounding diagram must fulfill the following condition:
Given a match consisting of exactly a set $X$ of abstract idioms, there must not exist an abstract idiom $y$, different from all $x \in X$, in the diagram, such that, $x_1 \prec_1 y \prec_2 x_1$ for $x_1,x_2 \in X$ and $\prec_1,\prec_2 \in \{\prec_\mathit{data},\prec_\mathit{IO}\} $.
A violation of this condition would introduce a cycle into the wiring with at least one wire from the newly introduced idiom to $y$ and one in the reverse direction.

In the rest of this section we will explore how these two transformations can be used to find different variations of abstract implementations that still preserve the original intent.

\subsection{Combining multiple idioms}

We have already seen in Section~\ref{sec:intro} that sometimes idiomatic code will realize two or more abstract idioms within a single program fragment.
For example, the summation of a sequence of values can be done on the fly, eliminating the construction of an intermediate list value.
\begin{lstlisting}
  let
    go i s
      | i == n = pure s
      | otherwise = do
        x <- readLn
        go (i+1) (s+x)
  xs <- go 0 0
\end{lstlisting}

We might be tempted to simply encode the requirements of this program fragment in a new abstract idiom.
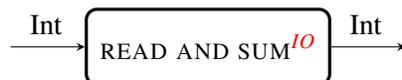
\begin{center}
  \begin{tikzpicture}
	\begin{pgfonlayer}{nodelayer}
		\node [style=Normal Box] (0) at (0, 0) {\idiomName{read and sum}\markSideEffect};
		\node [style=none] (1) at (-5.25, 0) {};
		\node [style=none] (2) at (5.25, 0) {};
		\node [style=Text node] (3) at (-5, 0.5) {Int};
		\node [style=Text node] (4) at (3.5, 0.5) {Int};
	\end{pgfonlayer}
	\begin{pgfonlayer}{edgelayer}
		\draw [style=Edge] (1.center) to (0);
		\draw [style=Edge] (0) to (2.center);
	\end{pgfonlayer}
\end{tikzpicture}
\end{center}

But this approach does not scale well.
Introducing a new abstract idiom for each combination of ``mergeable'' idioms goes against our goal of composable descriptions of intent.
For example, when computing the product of the list's values instead of the sum we would need a new abstract idiom but the concrete realizations of the loop-portion of that idiom are identical to the sum-case.

To facilitate better composition and reusability, we will instead define a new abstract idiom that accepts as parameters a starting value and a combination function, i.e., the parameters of a fold on the read list.
We also define a second abstract idiom providing these values.
Combining these two new idioms gives us an abstract implementation that can be instantiated to the Haskell code above.
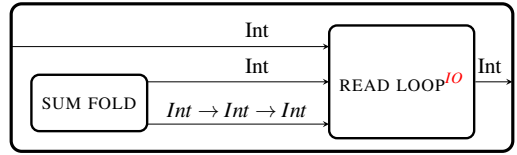
\begin{center}
  \insertDiagramWithBorder{outside-no-xsep}{diagrams/abstract-op/readLoopAlg}{}
\end{center}

Instead of a specialized summation idiom we now use an operation that provides the parameters to a list-fold and \idiomName{read loop} uses these parameters.
This abstract implementation has the same outside signature as \idiomName{read and sum} but models the data-flow on the inside in more detail.
When changing the summation to a product we can simply replace the abstract idiom providing the fold parameters and leave the loop idiom untouched.

But notice how the order of the abstract idiom representing the loop and the abstract idiom for the summation part has reversed in the data-flow for the new abstract implementation.
Previously information flowed from \idiomName{read list} into \idiomName{sum}, now information flows from \idiomName{sum fold} to \idiomName{read loop}.
Does this alter the intent underlying the abstract implementation?

Fortunately, the intent is still intact.
To see why, let us say that in the modified abstract implementation the idiom responsible for summation was moved ``upwards'' in the abstract program to a position before the reading-loop.
But it is not really the act of summing the list values that was moved before the loop but rather the information of how the summation can be achieved as a fold over a list.
In other words, the only information that has changed locations inside the program is static information and moving this static information around inside the abstract program does not change the overall behavior.

\subsection{Alternatives and contexts}
\label{sec:alternatives-and-contexts}

Now, how can we automatically find modular alternative abstract implementations, like the \idiomName{sum fold} and \idiomName{read loop} decomposition above, for a given abstract implementation?
Again, explicitly naming all combinations of abstract idioms that can be transformed into alternative data-flow patterns goes against our goal of a compositional and reusable framework.

We can separate the transformation into two stages.
First some abstract idioms are transformed into data-flow alternatives introducing contexts that can then potentially be merged into another idiom.
Next, we check for all abstract idioms in the abstract implementation whether they can consume one or more of the newly introduced contexts.
Both of these steps can be specified on a per-idiom basis eliminating the need to explicitly name combinations of idioms that can be merged.

For \idiomName{sum}, we get the following alternative version offering a list-fold context to its surroundings.
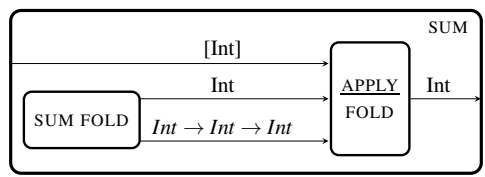
\begin{center}
  \insertDiagramWithBorder{outside-no-xsep, inner ysep=2pt}{diagrams/abstract-op/sumAlt}{}
\end{center}
In any abstract implementation this variant and the atomic \idiomName{sum} can be used interchangeably.
The substitution procedure defined above exactly captures such replacements.
We can think about this specific substitution as zooming into the internal structure of \idiomName{sum} exposing it to its surroundings.

The alternative form for summation introduces an additional abstract idiom \idiomApply{} \idiomName{fold}.
This idiom acts as a connector integrating the \idiomName{sum fold} idiom into the old data-flow.
As the name suggests, the intent of the \idiomApply{} idiom is a specialized function application.
From the types alone, \idiomApply{} \idiomName{fold} needs to produce an integer from a list of integers and the fold-parameters.
The natural way to achieve this is to simply apply a fold to the list.
Combined with \idiomName{sum fold} this could be realized as the following program fragment.
\begin{lstlisting}
  let (z,f) = (0,(+))
  let r = foldr z f xs
\end{lstlisting}
Which is a complicated (slightly non-idiomatic) way of writing the more straightforward version below.
\begin{lstlisting}
  let r = foldr 0 (+) xs
\end{lstlisting}

But the latter is a perfectly valid concrete idiom for the atomic \idiomName{sum} idiom we started with.
So in a sense the alternative implementation of the sum-idiom decomposes \idiomName{sum} into the application of a fold to its parameters.

\idiomApply{} idioms play a special role in our approach to artifact generation because they are not intended to actually be instantiated by concrete artifact fragments (even though in principle they could be, as we have just seen).
They do not represent new intent but result from splitting an atomic abstract idiom into a function application.
They are only the catalysts that drive the process of building specialized abstract idioms in a compositional manner and in the end will be consumed by another abstract idiom.
For that, we give for each abstract idiom a set of merging rules that allow for merging of different versions of \idiomApply{} idioms where applicable, eliminating the \idiomApply{} idioms in the process.
A merge rule consists of a pattern of abstract idioms, i.e., a well-formed abstract implementation, together with a name for the operation resulting from the merge.
We distinguish merge rules from alternative implementations by drawing a dashed line around the inner idioms, highlighting the new outside interface resulting from the merge.
Applying a merge rule to an abstract implementation is done through our previously defined merging procedure.
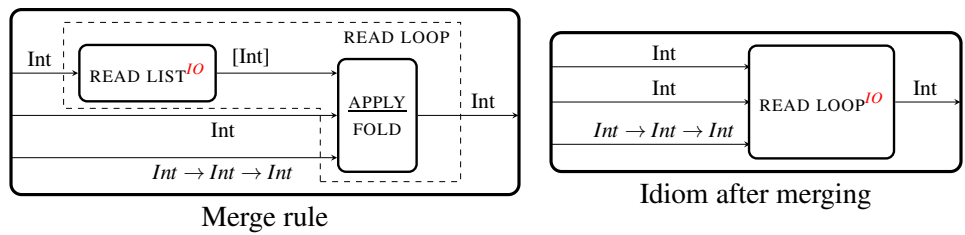
\begin{center}
  \insertDiagramWithBorder{outside-no-xsep}{diagrams/abstract-op/readListLhs}{Merge rule}
  \insertDiagramWithBorder{outside-no-xsep}{diagrams/abstract-op/readListRhs}{Idiom after merging}
\end{center}
Note, that in the above merging rule for the \idiomName{read list} idiom only the \idiomApply{} idiom for the fold parameters is part of the rule, making it independent of the fold-providing idiom connected to \idiomApply{} \idiomName{fold}.

In the same way we viewed alternative implementations as zooming into an idiom's internal structure, merge-rules can be viewed as zooming out again abstracting internal structure.
The fact that zooming in and out overlap only partially then leads to a variation of the abstract implementation.

For both alternative implementations and merge-rules we need to make sure that the transformations preserve intent.
That is, alternative implementations are specialized and more fine-grained versions of their respective base idioms, and the result of a merge-rule captures exactly the intent of the combined idioms.
Of course, as the labels of these new abstract idioms are purely syntactic, preservation of intent cannot be enforced or checked automatically at the level of abstract implementations.
Instead, we formulate requirements relating different concrete artifacts to each other making sure they are semantically coherent (see Section~\ref{sec:coherence}).
Creating new alternative implementations and merge-rules therefore must be done in accordance with these coherence conditions.

\subsection{Contexts with effects}

In Figure~\ref{fig:idiom-table} we have already seen another case of a context we can merge into our new \idiomName{read loop} idiom.
Once we merged the summation into the recursive loop function we can also do the same for printing the sum itself.
For this transformation we can define the alternative implementation and merging rule below.
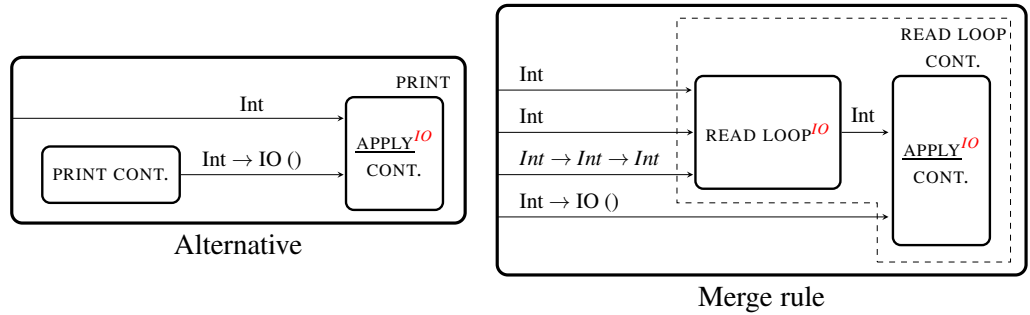
\begin{center}
  \insertDiagramWithBorder{outside-no-xsep}{diagrams/abstract-op/printAlt}{Alternative}
  \insertDiagramWithBorder{outside-no-xsep}{diagrams/abstract-op/printContLhs}{Merge rule}
\end{center}

Here the static context information is the continuation we need to apply to the sum and similarly the \idiomApply{} idiom's intent is application of this continuation to the result of the summation.
But now there is also an IO effect on the \idiomApply{} idiom.

It is important to note that the new abstract idioms introduced by alternative implementations and merge-rules do not necessarily have concrete counterparts for every possible kind of artifact.
While passing around continuations is a common pattern in Haskell programs, other kinds of artifacts, like our IO-specification language, do not have corresponding mechanisms with which such a pattern could be realized.
The variants of abstract implementations we get from alternatives and merge-rules can therefore be more specific to certain kinds of artifacts than the abstract implementation we derive them from.

\section{Artifact generation}
\label{sec:generation}

We have seen how we can describe the intent that underlies the different kinds of artifacts we want to generate through wiring diagrams connecting different abstract idioms.
Before we look at the concrete counterparts of these abstract idioms in the next section, we first describe how we can turn an intent description, i.e., an abstract implementation, into different kinds of artifacts.
For now all we need to know is that concrete idioms require inputs from other concrete idioms, create an artifact fragment of the targeted artifact kind, and produce additional outputs to be consumed by the other idioms.
For example, a concrete idiom for Haskell code for the \idiomName{read} operation will produce, in addition to Haskell code, as output the name of the variable the read value is bound to.
Likewise, a concrete idiom for \idiomName{print} requires an argument expression for the print function as input to produce Haskell code.

Given an abstract implementation and at least one concrete idiom of the desired artifact kind with matching signature per unique abstract idiom, we then choose for each abstract idiom a concrete idiom and resolve the data dependencies between them as defined by the wiring of the abstract implementation.
With a concrete idiom chosen for each abstract one, we now have an artifact fragment for each of them.
Next, we need to decide in which order these fragments should be combined.
For that, we can choose any total order that is compatible with both the order of IO-effects and the data-flow relation of the abstract idioms in the abstract implementation.
Then, we combine the corresponding concrete artifacts according to the chosen order into a single artifact.
For example, for a Haskell program we vertically glue code fragments together into a complete program.

We can repeat this process with different choices for concrete idioms to create multiple different artifacts for the same intent.
This can also mean using completely different concrete idioms to, for example, generate a different kind of artifact like a behavior specification instead of program code.

While differences in available concrete idioms already lead to variations in generated artifacts, we can explore even more variants using the techniques developed in the previous section.
The process of creating the final artifact remains the same, but we now explore different variants of the original abstract implementation first and then use the resulting data-flow variant as the input to the process described above.

The exploration of an abstract implementation's variants is separated into two stages.
These two stages correspond to the introduction and elimination of contexts described in Section~\ref{sec:alternatives-and-contexts}.

Given a set of alternative implementations, we first decide for each abstract idiom in the abstract implementation whether we want to replace it with a compatible alternative.
Using the substitution procedure we then insert the selected alternatives, possibly introducing \idiomApply{} idioms.
We then apply merge-rules to the wiring diagram until no further merging is possible.
If in the resulting diagram all contexts introduced by the substitutions are consumed again, i.e., if the diagram does not contain any \idiomApply{} idioms we have found an implementation variant.
Varying choices in the substitution step allows us to find more than a single variant of the original abstract implementation.
Each of the variants we obtain this way makes slightly different decisions regarding which context-sensitive data-flow variants to use.
Of course the abstract implementation we start with is itself always a possible result of this process.

Note that both the exploration of data-flow variants and the generation of a concrete artifact can be done randomly or exhaustively, but it is perfectly fine for both to pick a different mode.
It is possible and in fact can be desirable to use different combinations of these modes for different purposes.
For example, choosing a single abstract implementation variant at random but fully exploring its concrete instantiations yields several concrete implementations of the same problem decomposition.
On the other hand choosing a single random concrete instantiation for every possible variant of the initial abstract implementation can serve to highlight different possible solution strategies.

\subsection{Artifact coherence}
\label{sec:coherence}

While our intention is for artifacts to be realizations of a specific intent there is no formal connection between the wiring diagram of the abstract implementation and the artifact resulting from the generation procedure other than the generation process itself.
This is mainly because we have, deliberately, not defined formal semantics for abstract implementations.
Artifact generation interprets the purely syntactic abstract implementations into a concrete artifact.
A fixed set of concrete idioms, alternative implementations, and merge-rules thereby gives semantics to abstract implementations in the form of sets of concrete artifacts resulting from the generation process.
For such an interpretation to be well-defined, all possible artifacts that can be generated from an abstract implementation or its variants must be semantically coherent.
Otherwise, we cannot use artifacts of the same kind interchangeably and multiple different artifacts might contradict each other instead of supplementing each other.

The generation process does not do anything to check coherence but assumes that the set of concrete idioms is built such that possible results are coherent by construction.
Validation of concrete idiom sets, alternatives, and merge-rules is therefore the responsibility of the educator but can sometimes be partially automated.
For example, when generating multiple different programs, artifact coherence requires all programs to be pairwise equivalent in their observable IO-behavior.
If we additionally generate specification expressions these also must be equivalent among themselves and also all generated programs must satisfy all specifications.
The latter property can be tested automatically with our previously presented testing framework~\cite{tfpie19-paper}.

The definition of coherence can also vary between different applications.
For example, we might treat programs as having equivalent behavior if they only differ in the way they embellish a printed result or how they prompt for inputs.
This flexibility to choose coherence conditions based on application context would be lost if we were to define global semantics for abstract implementations.

\section{Concrete idioms}
\label{sec:concrete}

The quality and especially the diversity of generated artifacts critically depends on the concrete idioms available.
This section will go through the process of defining various concrete idioms to build Haskell programs for the following abstract implementation.
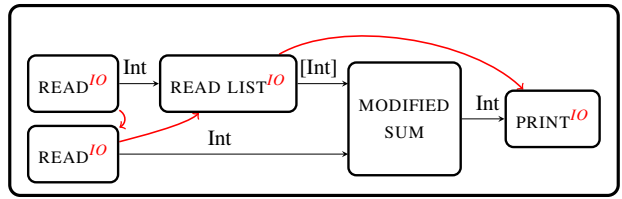
\begin{center}
  \insertDiagramWithBorder{outside}{diagrams/abstract-op/msum-example}{}
\end{center}

By doing so we will introduce various different forms of concrete idioms.
Concrete idioms are given in a graphical notation but are in essence functions producing artifact fragments.
We will also see that it is useful to sometimes equip concrete idioms, or more precisely the data produced by them, with some form of higher-order outputs.

\subsection{Basic idioms}

Each concrete idiom is in essence a function that takes inputs from other concrete idioms and produces an artifact fragment together with additional outputs for other idioms to use.
Take, for example, the abstract idiom below.
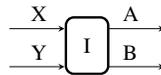
\begin{center}
  \insertDiagramWithBorder{}{diagrams/abstract-op/xy2ab}{}
\end{center}

A corresponding concrete idiom for artifact kind $k$ is a function $F_k(X) \times F_k(Y) \rightarrow M_k \times F_k(A) \times F_k(B)$.
Here $M_k$ is the set of artifact fragments for the kind of artifact we want to generate, e.g., program code, specifications, etc.
We require $M_k$ to be a monoid.
$F_k$ maps abstract types to sets, interpreting the abstract types based on the choice of $k$.
The artifact fragments resulting from the evaluation of the concrete idioms for an abstract implementation, i.e., values from $M_k$, are combined using $M_k$'s monoidal operation and sequenced by a total order compatible with the abstract implementation's data-flow and IO-effect ordering.

At the most basic level a concrete idiom is simply a parameterized artifact fragment that provides parts of this fragment as additional outputs.
Take as an example two concrete idioms for the abstract \idiomName{read~list} idiom:
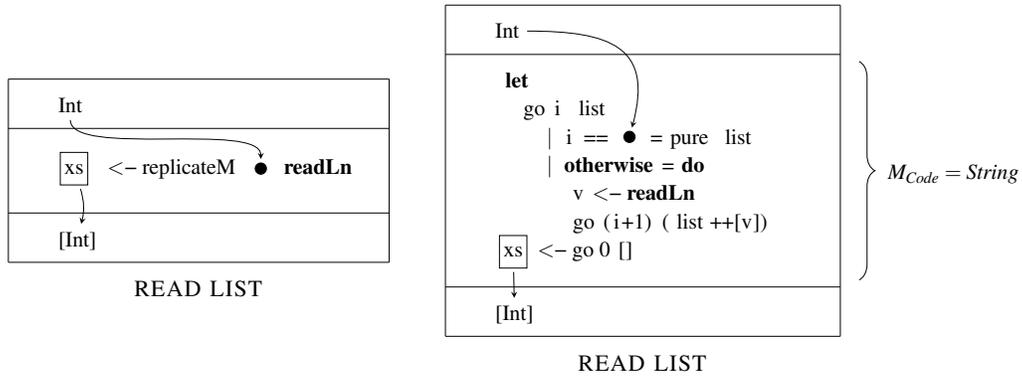
\begin{center}
  \insertDiagramWithBorder{}{diagrams/idioms/programs/readN}{\idiomName{read list}}
  \insertDiagramWithBorder{}{diagrams/idioms/programs/readNLoop}{\idiomName{read list}}
  \hspace{-5mm}\insertDiagramWithBorder{}{diagrams/idioms/programs/brace}{}
\end{center}

The graphical notation of concrete idioms is a box with three subdivisions.
The first containing the input types to this idiom and the last containing the types of the additional outputs
($F_\mathit{Code}$ here maps to sets of well-typed expressions of the respective Haskell types).
The middle part contains the artifact fragment produced by the idiom.
Parts of the fragment filled in by information from other idioms are represented as \tikz \node[inputDot]{};.
The parts of the fragment that are provided as additional outputs are marked with boxes.
Both inputs and outputs are connected with arrows to their respective input and output positions.

For simplicity all concrete idioms shown here will use fixed names for program variables.
In practice these names are drawn from a supply of fresh names to avoid name clashes.

\subsection{Idioms without visible output}

Not all concrete idioms need to produce visible artifacts.
Sometimes we just want them to create an expression to be used as input to another idiom.
For example, for the \idiomName{modified sum} idiom we can either bind the result using a \texttt{let}-statement or simply pass on an expression.
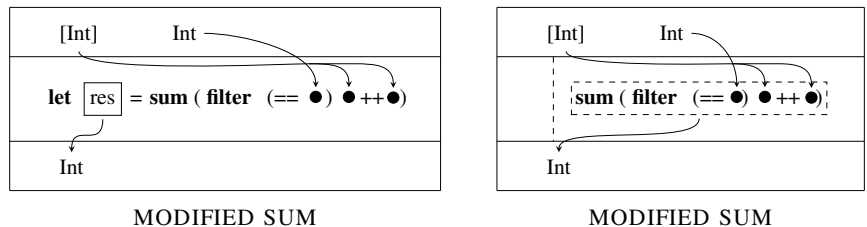
\begin{center}
  \insertDiagramWithBorder{}{diagrams/idioms/programs/msum1}{\idiomName{modified sum}}
  \insertDiagramWithBorder{}{diagrams/idioms/programs/msum2}{\idiomName{modified sum}}
\end{center}

The second idiom has no visible result artifact.
From a formal perspective, we can think of it as producing the neutral element of the underlying artifact monoid $M_k$, which here is the empty string.
We separate the middle part of the idiom into two parts: one for the, possibly empty, artifact output and one for the additional silent outputs.
While we often have either visible artifacts or silent outputs, it is perfectly fine for concrete idioms to have both simultaneously (see Section~\ref{sec:other} for examples).

In our example abstract implementation the output of \idiomName{modified sum} is passed into \idiomName{print}.
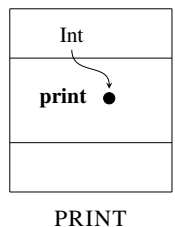
\begin{center}
  \insertDiagramWithBorder{}{diagrams/idioms/programs/print}{\idiomName{print}}
\end{center}

In case the expression for our specialized summation was not bound with \texttt{let} before it is then inserted into the produced artifact by the concrete \idiomName{print} idiom here.

For \idiomName{print}, we have also already seen an alternative implementation
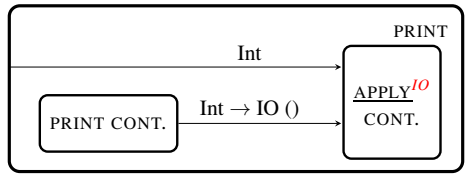
\begin{center}
  \insertDiagramWithBorder{outside-no-xsep}{diagrams/abstract-op/printAlt}{}
\end{center}

\idiomName{print cont.} is another example of a concrete idiom that in most cases will not produce visible output (see left idiom below).
A variant with visible output is, in principle, possible, but will produce less idiomatic code (e.g.\ the right idiom below).
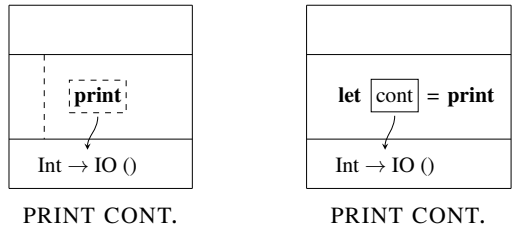
\begin{center}
  \insertDiagramWithBorder{}{diagrams/idioms/programs/printCont}{\idiomName{print cont.}}
  \qquad
  \insertDiagramWithBorder{}{diagrams/idioms/programs/printContVis}{\idiomName{print cont.}}
\end{center}

\subsection{``Higher-order'' idioms}
It is often natural to have concrete idioms produce incomplete expressions to pass to the next idiom.
This is especially the case in higher-order scenarios but can also occur in pseudo higher-order situations when concrete idioms produce syntactic contexts, e.g., an if-then-else construct, surrounding one or more values in a ``downstream'' idiom.
In essence, we want concrete idioms to produce outputs that still have holes in them and then fill these holes with information from other idioms to which this output is passed.

As motivation, let us look at an alternative implementation for \idiomName{modified sum} that we have already hinted at in the discussion of merge-rule applicability.
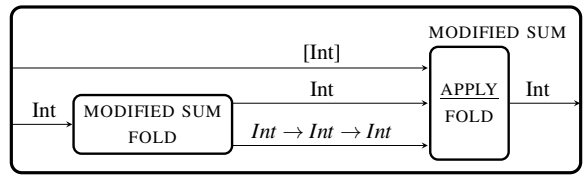
\begin{center}
  \insertDiagramWithBorder{outside-no-xsep}{diagrams/abstract-op/msumAltWrong}{}
\end{center}

At first glance this seems fine, and looking only at the abstract implementations it is, but problems arise when we try to define concrete idioms for \idiomName{modified sum}.
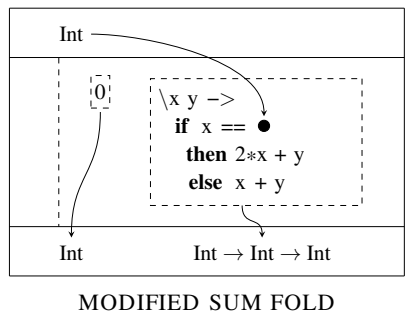
\begin{center}
  \insertDiagramWithBorder{}{diagrams/idioms/programs/msumWrong}{\idiomName{modified sum fold}}
\end{center}

We now run into a problem after defining the merge-rule for \idiomName{read list} and defining a concrete idiom for the resulting \idiomName{read loop}.
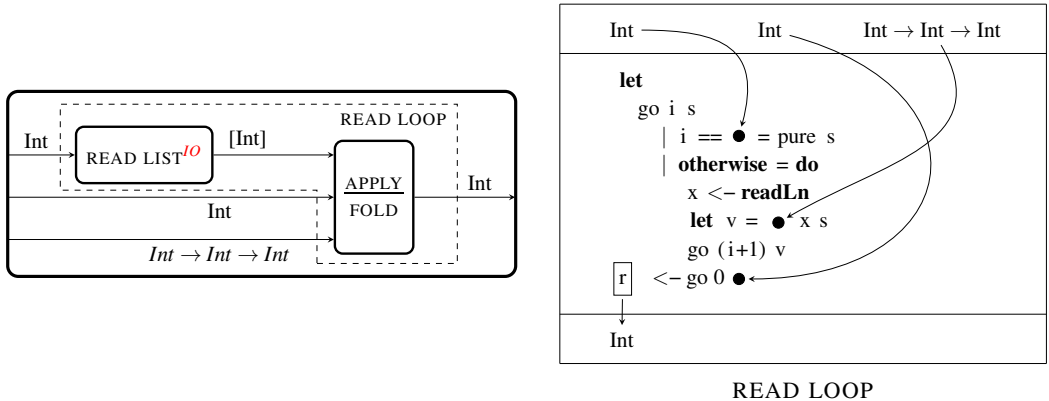
\begin{center}
  \insertDiagramWithBorder{outside-no-xsep}{diagrams/abstract-op/readListLhs}{}
  \insertDiagramWithBorder{}{diagrams/idioms/programs/readLoopWrong}{\idiomName{read loop}}
\end{center}

Using the output of \idiomName{modified sum} in this concrete idiom for \idiomName{read loop} produces the expression \lstinline|(\x y -> if ...)x s| in the idiom's artifact.
This is highly non-idiomatic as this expression can immediately be $\beta$-reduced.

The problem here is that the function argument to \idiomName{read loop} must be a complete expression when in actuality it should use information not known inside \idiomName{modified sum}, here the names \lstinline|x| and \lstinline|s|.
This simple structure of passing function arguments between concrete idioms only works for point-free expressions, like outputting \lstinline|(+)| in \idiomName{sum fold} or \lstinline|print| in \idiomName{print cont.}.
In the general case, however, it is insufficient.
We therefore introduce the ability to pass along incomplete outputs to other concrete idioms where their missing arguments get filled in.
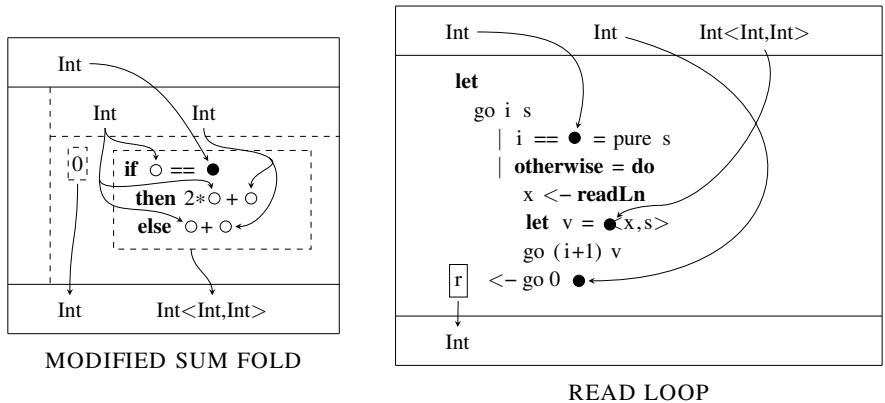
\begin{center}
  \insertDiagramWithBorder{}{diagrams/idioms/programs/msumAlg}{\idiomName{modified sum fold}}
  \insertDiagramWithBorder{}{diagrams/idioms/programs/readLoop}{\idiomName{read loop}}
\end{center}

It is important that such partial outputs of concrete idioms are only part of silent outputs.
They provide artifact fragments to be completed with ``downstream'' data.
Allowing such fragments as part of the visible result of an idiom would introduce unwanted cyclic dependencies within the resulting artifact.

Equipped with these new tools we can now define the improved versions of \idiomName{modified sum}'s alternative implementation and the corresponding merge-rule.

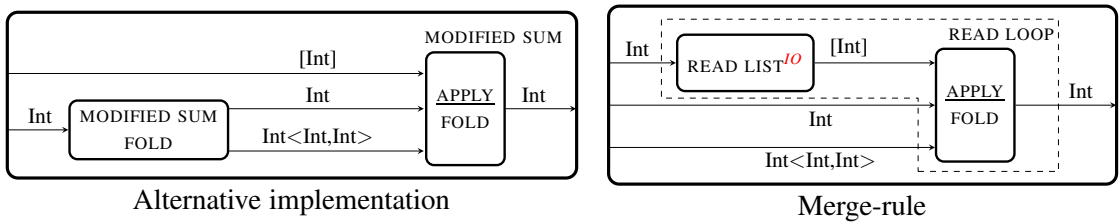
\begin{center}
  \insertDiagramWithBorder{outside-no-xsep}{diagrams/abstract-op/msumAlt}{Alternative implementation}
  \insertDiagramWithBorder{outside-no-xsep}{diagrams/abstract-op/readLoopMerge}{Merge-rule}
\end{center}

When generating Haskell program code this way of handling these higher-order arguments may seem unnecessarily complicated.
After all we could just generate programs containing non-idiomatic applications of lambda-expressions and then clean up the code in a post-processing step.
While this would work for Haskell programs, not all artifact kinds have access to such $\beta$-reduction-like post-processing steps.
For example, we cannot easily do such a post-processing step for our specifications of IO-behavior or for natural language descriptions.

\subsection{Generating concrete artifacts}

Now that we have defined concrete idioms for the abstract idioms in our example we can finally generate idiomatic programs for our abstract implementation.

As already mentioned in Section~\ref{sec:generation} we find a variant of the abstract implementation for which we want to generate programs.
Next we choose a specific concrete idiom for each occurrence of an abstract idiom.
Then the outputs are propagated to inputs according to the wiring structure of the abstract implementation.
After choosing a total order on the boxes of the abstract implementation that is compatible with the data-flow and the order of the IO-effects we combine the individual concrete program fragments into a single program according to the chosen order.
For programs this means simply concatenating the results vertically.

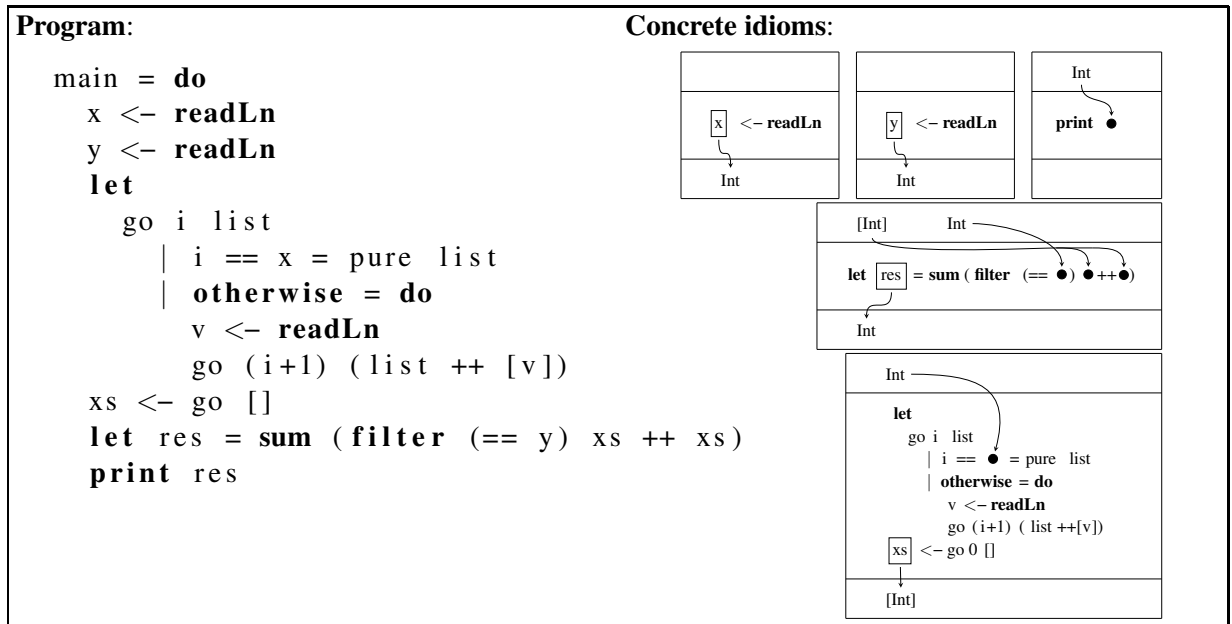
\begin{figure}
  \begin{minipage}[t]{.5\textwidth}
    \textbf{Program}:
    \begin{lstlisting}[language=haskell]
  main = do
    x <- readLn
    y <- readLn
    let
      go i list
        | i == x = pure list
        | otherwise = do
          v <- readLn
          go (i+1) (list ++ [v])
    xs <- go []
    let res = sum (filter (== y) xs ++ xs)
    print res
    \end{lstlisting}
  \end{minipage}
  \begin{minipage}[t]{.45\textwidth}
    \textbf{Concrete idioms}:
    \begin{flushright}
      \scalebox{.6}{\begin{tikzpicture}
	\begin{pgfonlayer}{nodelayer}
		\node [style=Text node] (0) at (-3.175, 0.075) {\lstinline{<- readLn}};
		\node [style=Output] (1) at (-4.3, 0.075) {\lstinline{x}};
		\node [style=Text node] (6) at (-4, -2.5) {Int};
		\node [style=none] (7) at (-5.5, 1.5) {};
		\node [style=none] (8) at (1.5, 1.5) {};
		\node [style=none] (9) at (-5.5, -1.5) {};
		\node [style=none] (10) at (1.5, -1.5) {};
		\node [style=none] (11) at (-5.5, 3.25) {};
		\node [style=none] (12) at (1.5, 3.25) {};
		\node [style=none] (13) at (1.5, -3.25) {};
		\node [style=none] (14) at (-5.5, -3.25) {};
	\end{pgfonlayer}
	\begin{pgfonlayer}{edgelayer}
		\draw [style=Edge, in=90, out=-90, looseness=2.25] (1) to (6);
		\draw (9.center) to (10.center);
		\draw (7.center) to (8.center);
		\draw (11.center) to (12.center);
		\draw (12.center) to (13.center);
		\draw (13.center) to (14.center);
		\draw (14.center) to (11.center);
	\end{pgfonlayer}
\end{tikzpicture}}
      \scalebox{.6}{\begin{tikzpicture}
	\begin{pgfonlayer}{nodelayer}
		\node [style=Text node] (0) at (-3.175, 0.075) {\lstinline{<- readLn}};
		\node [style=Output] (1) at (-4.3, 0.075) {\lstinline{y}};
		\node [style=Text node] (6) at (-4, -2.5) {Int};
		\node [style=none] (7) at (-5.5, 1.5) {};
		\node [style=none] (8) at (1.5, 1.5) {};
		\node [style=none] (9) at (-5.5, -1.5) {};
		\node [style=none] (10) at (1.5, -1.5) {};
		\node [style=none] (11) at (-5.5, 3.25) {};
		\node [style=none] (12) at (1.5, 3.25) {};
		\node [style=none] (13) at (1.5, -3.25) {};
		\node [style=none] (14) at (-5.5, -3.25) {};
	\end{pgfonlayer}
	\begin{pgfonlayer}{edgelayer}
		\draw [style=Edge, in=90, out=-90, looseness=2.25] (1) to (6);
		\draw (9.center) to (10.center);
		\draw (7.center) to (8.center);
		\draw (11.center) to (12.center);
		\draw (12.center) to (13.center);
		\draw (13.center) to (14.center);
		\draw (14.center) to (11.center);
	\end{pgfonlayer}
\end{tikzpicture}}
      \scalebox{.6}{\begin{tikzpicture}
	\begin{pgfonlayer}{nodelayer}
		\node [style=Text node] (0) at (-4.75, 0.075) {\lstinline{print}};
		\node [style=Input Dot] (3) at (-1.975, 0.075) {};
		\node [style=Text node] (5) at (-4, 2.25) {Int};
		\node [style=none] (7) at (-5.5, 1.5) {};
		\node [style=none] (8) at (0.25, 1.5) {};
		\node [style=none] (9) at (-5.5, -1.5) {};
		\node [style=none] (10) at (0.25, -1.5) {};
		\node [style=none] (11) at (-5.5, 3.25) {};
		\node [style=none] (12) at (0.25, 3.25) {};
		\node [style=none] (13) at (0.25, -3.25) {};
		\node [style=none] (14) at (-5.5, -3.25) {};
	\end{pgfonlayer}
	\begin{pgfonlayer}{edgelayer}
		\draw [style=Edge, in=90, out=-90, looseness=1.25] (5) to (3);
		\draw (9.center) to (10.center);
		\draw (7.center) to (8.center);
		\draw (11.center) to (12.center);
		\draw (12.center) to (13.center);
		\draw (13.center) to (14.center);
		\draw (14.center) to (11.center);
	\end{pgfonlayer}
\end{tikzpicture}}

      \scalebox{.6}{\begin{tikzpicture}
	\begin{pgfonlayer}{nodelayer}
		\node [style=Text node] (1) at (-1.5, 0) {\lstinline[mathescape]{= sum (filter (==  $\;\;$)  $\;\;$ ++ $\;\;$)}};
		\node [style=Text node] (5) at (-4, 2.25) {[Int]};
		\node [style=Text node] (6) at (-4, -2.5) {Int};
		\node [style=none] (7) at (-5.5, 1.5) {};
		\node [style=none] (8) at (9.775, 1.5) {};
		\node [style=none] (9) at (-5.5, -1.5) {};
		\node [style=none] (10) at (9.775, -1.5) {};
		\node [style=none] (11) at (-5.5, 3.25) {};
		\node [style=none] (12) at (9.775, 3.25) {};
		\node [style=none] (13) at (9.775, -3.25) {};
		\node [style=none] (14) at (-5.5, -3.25) {};
		\node [style=Output] (15) at (-3, 0) {\lstinline{res}};
		\node [style=Text node] (16) at (-4.5, 0) {\lstinline{let}};
		\node [style=Text node] (17) at (0, 2.25) {Int};
		\node [style=Input Dot] (18) at (5.35, 0.075) {};
		\node [style=Input Dot] (21) at (6.525, 0.075) {};
		\node [style=Input Dot] (22) at (8.1, 0.05) {};
		\node [style=none] (23) at (5.1, 1.25) {};
		\node [style=none] (24) at (6.7, 1.25) {};
	\end{pgfonlayer}
	\begin{pgfonlayer}{edgelayer}
		\draw (9.center) to (10.center);
		\draw (7.center) to (8.center);
		\draw (11.center) to (12.center);
		\draw (12.center) to (13.center);
		\draw (13.center) to (14.center);
		\draw (14.center) to (11.center);
		\draw [style=Edge, in=90, out=-90, looseness=1.75] (15) to (6);
		\draw [style=Edge, in=90, out=0, looseness=0.75] (17) to (18);
		\draw (23.center) to (24.center);
		\draw [style=Edge, in=90, out=0, looseness=1.25] (23.center) to (21);
		\draw [style=Edge, in=90, out=0, looseness=1.25] (24.center) to (22);
		\draw [in=-180, out=-90, looseness=0.25] (5) to (23.center);
	\end{pgfonlayer}
\end{tikzpicture}}

      \hfill\scalebox{.6}{\begin{tikzpicture}
	\begin{pgfonlayer}{nodelayer}
		\node [style=Text node] (0) at (-4.75, 4) {\lstinline{let}};
		\node [style=Output] (1) at (-4.725, -2) {\lstinline{xs}};
		\node [style=Input Dot] (3) at (0.025, 2.075) {};
		\node [style=Text node, text centered] (4) at (-4, 3) {\lstinline{go i list}};
		\node [style=Text node] (5) at (-5, 5.75) {Int};
		\node [style=Text node] (6) at (-5, -4.25) {[Int]};
		\node [style=none] (7) at (-6.5, 5) {};
		\node [style=none] (8) at (7.5, 5) {};
		\node [style=none] (9) at (-6.5, -3.25) {};
		\node [style=none] (10) at (7.5, -3.25) {};
		\node [style=none] (11) at (-6.5, 6.75) {};
		\node [style=none] (12) at (7.5, 6.75) {};
		\node [style=none] (13) at (7.5, -5) {};
		\node [style=none] (14) at (-6.5, -5) {};
		\node [style=Text node] (15) at (-3.25, 2) {\lstinline{| i ==}};
		\node [style=Text node] (16) at (-3.25, 1) {\lstinline{| otherwise = do}};
		\node [style=Text node] (18) at (-2.25, 0) {\lstinline{v <- readLn}};
		\node [style=Text node] (19) at (-2.25, -1) {\lstinline{go (i+1) (list++[v])}};
		\node [style=Text node] (20) at (-3.5, -2) {\lstinline{<- go 0 []}};
		\node [style=Text node] (21) at (0.5, 2) {\lstinline{= pure list}};
	\end{pgfonlayer}
	\begin{pgfonlayer}{edgelayer}
		\draw [style=Edge, in=75, out=0, looseness=1.50] (5) to (3);
		\draw [style=Edge] (1) to (6);
		\draw (9.center) to (10.center);
		\draw (7.center) to (8.center);
		\draw (11.center) to (12.center);
		\draw (12.center) to (13.center);
		\draw (13.center) to (14.center);
		\draw (14.center) to (11.center);
	\end{pgfonlayer}
\end{tikzpicture}}
    \end{flushright}
  \end{minipage}
  \caption{\label{fig:result1} Program built from concrete idioms}
\end{figure}

\begin{figure}
  \textbf{Concrete idioms}:
  \begin{center}
    \scalebox{.6}{\begin{tikzpicture}
	\begin{pgfonlayer}{nodelayer}
		\node [style=Text node] (0) at (-3.175, 0.075) {\lstinline{<- readLn}};
		\node [style=Output] (1) at (-4.3, 0.075) {\lstinline{x}};
		\node [style=Text node] (6) at (-4, -2.5) {Int};
		\node [style=none] (7) at (-5.5, 1.5) {};
		\node [style=none] (8) at (1.5, 1.5) {};
		\node [style=none] (9) at (-5.5, -1.5) {};
		\node [style=none] (10) at (1.5, -1.5) {};
		\node [style=none] (11) at (-5.5, 3.25) {};
		\node [style=none] (12) at (1.5, 3.25) {};
		\node [style=none] (13) at (1.5, -3.25) {};
		\node [style=none] (14) at (-5.5, -3.25) {};
	\end{pgfonlayer}
	\begin{pgfonlayer}{edgelayer}
		\draw [style=Edge, in=90, out=-90, looseness=2.25] (1) to (6);
		\draw (9.center) to (10.center);
		\draw (7.center) to (8.center);
		\draw (11.center) to (12.center);
		\draw (12.center) to (13.center);
		\draw (13.center) to (14.center);
		\draw (14.center) to (11.center);
	\end{pgfonlayer}
\end{tikzpicture}}
    \scalebox{.6}{\begin{tikzpicture}
	\begin{pgfonlayer}{nodelayer}
		\node [style=Text node] (0) at (-3.175, 0.075) {\lstinline{<- readLn}};
		\node [style=Output] (1) at (-4.3, 0.075) {\lstinline{y}};
		\node [style=Text node] (6) at (-4, -2.5) {Int};
		\node [style=none] (7) at (-5.5, 1.5) {};
		\node [style=none] (8) at (1.5, 1.5) {};
		\node [style=none] (9) at (-5.5, -1.5) {};
		\node [style=none] (10) at (1.5, -1.5) {};
		\node [style=none] (11) at (-5.5, 3.25) {};
		\node [style=none] (12) at (1.5, 3.25) {};
		\node [style=none] (13) at (1.5, -3.25) {};
		\node [style=none] (14) at (-5.5, -3.25) {};
	\end{pgfonlayer}
	\begin{pgfonlayer}{edgelayer}
		\draw [style=Edge, in=90, out=-90, looseness=2.25] (1) to (6);
		\draw (9.center) to (10.center);
		\draw (7.center) to (8.center);
		\draw (11.center) to (12.center);
		\draw (12.center) to (13.center);
		\draw (13.center) to (14.center);
		\draw (14.center) to (11.center);
	\end{pgfonlayer}
\end{tikzpicture}}
    \scalebox{.6}{\begin{tikzpicture}
	\begin{pgfonlayer}{nodelayer}
		\node [style=Text node] (0) at (-2.25, 0.075) {\lstinline{<- replicateM}};
		\node [style=Output] (1) at (-3.8, 0.075) {\lstinline{xs}};
		\node [style=Input Dot] (3) at (3.45, 0.075) {};
		\node [style=Text node, text centered] (4) at (4, 0.075) {\lstinline{readLn}};
		\node [style=Text node] (5) at (-4, 2.25) {Int};
		\node [style=Text node] (6) at (-4, -2.5) {[Int]};
		\node [style=none] (7) at (-5.5, 1.5) {};
		\node [style=none] (8) at (8, 1.5) {};
		\node [style=none] (9) at (-5.5, -1.5) {};
		\node [style=none] (10) at (8, -1.5) {};
		\node [style=none] (11) at (-5.5, 3.25) {};
		\node [style=none] (12) at (8, 3.25) {};
		\node [style=none] (13) at (8, -3.25) {};
		\node [style=none] (14) at (-5.5, -3.25) {};
	\end{pgfonlayer}
	\begin{pgfonlayer}{edgelayer}
		\draw [style=Edge, in=90, out=-90, looseness=0.50] (5) to (3);
		\draw [style=Edge, bend left=15] (1) to (6);
		\draw (9.center) to (10.center);
		\draw (7.center) to (8.center);
		\draw (11.center) to (12.center);
		\draw (12.center) to (13.center);
		\draw (13.center) to (14.center);
		\draw (14.center) to (11.center);
	\end{pgfonlayer}
\end{tikzpicture}}
    \scalebox{.6}{\begin{tikzpicture}
	\begin{pgfonlayer}{nodelayer}
		\node [style=none] (16) at (-6.5, 1.5) {};
		\node [style=none] (17) at (-6.5, -1.5) {};
		\node [style=Silent Output] (20) at (-6, 0) {\lstinline[mathescape]{sum (filter (== $\;\;$)  $\;\;$ ++ $\;\;$)}};
		\node [style=Text node] (22) at (-7, 2.25) {[Int]};
		\node [style=Text node] (23) at (-7, -2.5) {Int};
		\node [style=none] (24) at (-8.5, 1.5) {};
		\node [style=none] (25) at (4.525, 1.5) {};
		\node [style=none] (26) at (-8.5, -1.5) {};
		\node [style=none] (27) at (4.525, -1.5) {};
		\node [style=none] (28) at (-8.5, 3.25) {};
		\node [style=none] (29) at (4.525, 3.25) {};
		\node [style=none] (30) at (4.525, -3.25) {};
		\node [style=none] (31) at (-8.5, -3.25) {};
		\node [style=Text node] (34) at (-3, 2.25) {Int};
		\node [style=Input Dot] (38) at (1, 0.075) {};
		\node [style=Input Dot] (39) at (0, 0.075) {};
		\node [style=Input Dot] (40) at (2.65, 0.05) {};
		\node [style=none] (41) at (-0.175, 1.25) {};
		\node [style=none] (42) at (1.5, 1.25) {};
	\end{pgfonlayer}
	\begin{pgfonlayer}{edgelayer}
		\draw [style=Dashed Edge] (16.center) to (17.center);
		\draw (26.center) to (27.center);
		\draw (24.center) to (25.center);
		\draw (28.center) to (29.center);
		\draw (29.center) to (30.center);
		\draw (30.center) to (31.center);
		\draw (31.center) to (28.center);
		\draw [style=Edge, in=90, out=-90, looseness=0.50] (20) to (23);
		\draw [in=-180, out=-90, looseness=0.25] (22) to (41.center);
		\draw [style=Edge, in=90, out=0] (34) to (39);
		\draw (41.center) to (42.center);
		\draw [style=Edge, in=90, out=0, looseness=1.25] (41.center) to (38);
		\draw [style=Edge, in=90, out=0, looseness=1.25] (42.center) to (40);
	\end{pgfonlayer}
\end{tikzpicture}}
    \scalebox{.6}{\begin{tikzpicture}
	\begin{pgfonlayer}{nodelayer}
		\node [style=Text node] (0) at (-4.75, 0.075) {\lstinline{print}};
		\node [style=Input Dot] (3) at (-1.975, 0.075) {};
		\node [style=Text node] (5) at (-4, 2.25) {Int};
		\node [style=none] (7) at (-5.5, 1.5) {};
		\node [style=none] (8) at (0.25, 1.5) {};
		\node [style=none] (9) at (-5.5, -1.5) {};
		\node [style=none] (10) at (0.25, -1.5) {};
		\node [style=none] (11) at (-5.5, 3.25) {};
		\node [style=none] (12) at (0.25, 3.25) {};
		\node [style=none] (13) at (0.25, -3.25) {};
		\node [style=none] (14) at (-5.5, -3.25) {};
	\end{pgfonlayer}
	\begin{pgfonlayer}{edgelayer}
		\draw [style=Edge, in=90, out=-90, looseness=1.25] (5) to (3);
		\draw (9.center) to (10.center);
		\draw (7.center) to (8.center);
		\draw (11.center) to (12.center);
		\draw (12.center) to (13.center);
		\draw (13.center) to (14.center);
		\draw (14.center) to (11.center);
	\end{pgfonlayer}
\end{tikzpicture}}
  \end{center}

  \textbf{Program}:
  \begin{lstlisting}[mathescape=false]
  main = do
    x <- readLn
    y <- readLn
    xs <- replicateM x readLn
    print $ sum (filter (== y) xs ++ xs)
  \end{lstlisting}
  \caption{\label{fig:result2} Another concrete program artifact}
\end{figure}
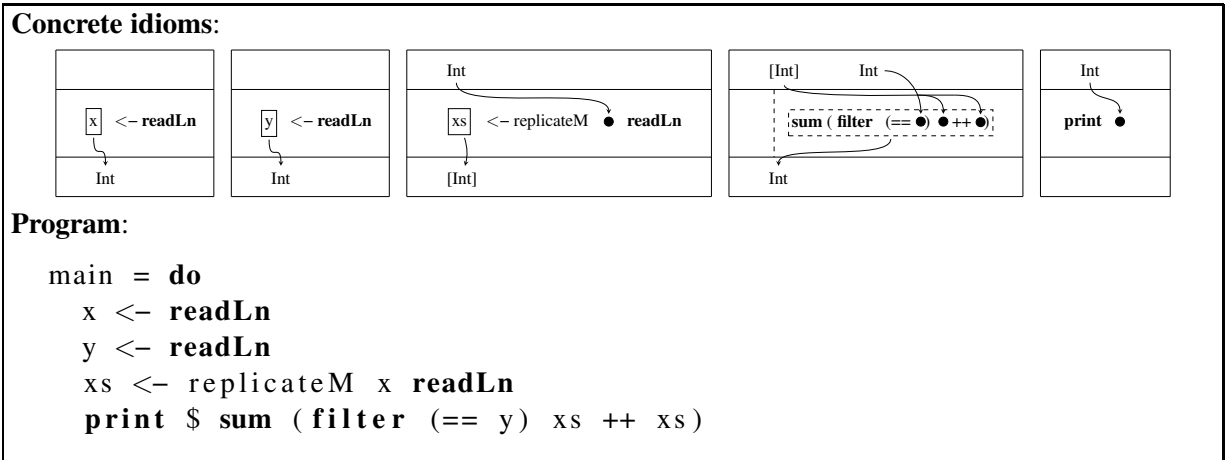

Figures~\ref{fig:result1} and \ref{fig:result2} show results for different choices of concrete idioms for the abstract implementation from the top of this section.
The more different concrete idioms we define for each abstract idiom the more different idiomatic programs can be generated.
Using other abstract implementations derived with the help of alternative implementations for abstract idioms and the corresponding merge-rules adds additional variations.
Given, for example, one concrete idiom for \idiomName{read} and \idiomName{print}, as well as for the variants \idiomName{read loop} and \idiomName{read loop cont.} together with 2 variants for \idiomName{read list}, 6 variants for \idiomName{modified sum}, and 2 for \idiomName{modified sum fold} we get a total of 20 slightly different concrete programs.
Adding more concrete idioms can quickly increase the total number of programs.
For example, adding a new concrete idiom for reading the list induces a new program variant for every compatible expression of the summation and vice versa.
Therefore, the more concrete idioms we already have for other abstract idioms the more new variants result from adding a single new concrete idiom for some abstract idiom.

\subsection{Idioms for other kinds of artifact}
\label{sec:other}

Instead of mapping abstract idioms to program fragments we can also map to specification expression fragments.
This way we can build specifications of behavior like the one in Figure~\ref{fig:example-with-spec}:

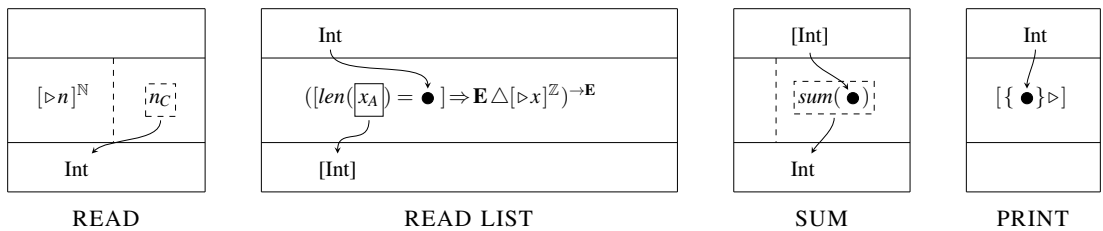
\begin{center}
  \insertDiagramWithBorder{}{diagrams/idioms/specifications/read}{\idiomName{read}}
  \insertDiagramWithBorder{}{diagrams/idioms/specifications/readN}{\idiomName{read list}}
  \insertDiagramWithBorder{}{diagrams/idioms/specifications/sum}{\idiomName{sum}}
  \insertDiagramWithBorder{}{diagrams/idioms/specifications/print}{\idiomName{print}}
\end{center}

\bigskip
\noindent
In the same way we can also create simple natural language descriptions of intended behavior.

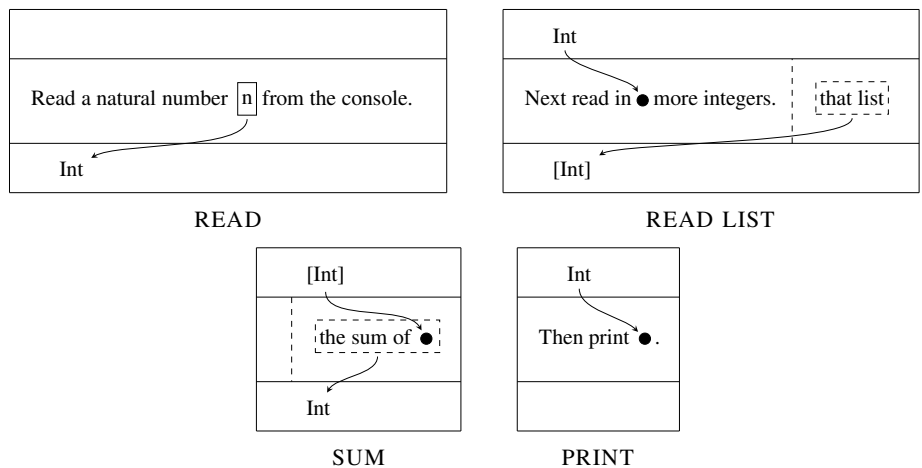
\begin{center}
  \insertDiagramWithBorder{}{diagrams/idioms/natural/read}{\idiomName{read}}
  \insertDiagramWithBorder{}{diagrams/idioms/natural/readN}{\idiomName{read list}}

  \insertDiagramWithBorder{}{diagrams/idioms/natural/sum}{\idiomName{sum}}
  \insertDiagramWithBorder{}{diagrams/idioms/natural/print}{\idiomName{print}}
\end{center}

In general our framework allows us to generate varying sets of corresponding artifacts, e.g., a specification expression together with a verbal description (task description) and a program (sample solution).

\section{\metaNameCap{}s}
\label{sec:meta}

So far we have seen how we can generate multiple idiomatic programs or related artifacts from a single description of behavior.
Recall that abstract implementations encode exactly one specific intent.
But for automatically generating exercise tasks it is often useful and desirable to also vary the underlying intent.
For example, we might want behavior from a family of related intents like first creating a list in some way and then consuming that list.

Conveniently, we can solve this problem easily with the tools introduced so far.
The same diagrammatic description we use for abstract implementations can also be used to describe a more general outline of intent.
That is, we connect boxes with wires where boxes can now also stand for broader classes of behavior instead of a single specific intent.
We call these diagrams \metaName{}s and in them denote boxes representing broader classes of behavior by a dashed border.

We can then define alternative forms for the dashed boxes in the same way we did for abstract idioms.
These alternatives are again \metaName{}s, i.e., built from a combination of abstract idioms and boxes for behavior classes.
This way we can define a system of refinement rules that produce abstract implementations similar to how a context-free grammar produces words of some formal language.
The dashed boxes act as the non-terminal symbols in these systems.
Figure~\ref{fig:grammar} shows an excerpt of such a grammar-like system.
The dashed red lines indicate how the potential side effects, introduced in the refinements of the dashed boxes, relate to the overall effect ordering.

Starting from an initial \metaName{} we can, randomly or exhaustively, explore different possible abstract implementations by applying substitutions until all boxes with dashed borders have been replaced and only abstract idioms remain.
We are then left with an abstract implementation to generate artifacts from.
Depending on the application scenario we can now seed the generation of exercise task artifacts either with an abstract implementation or an initial \metaName.

\begin{figure}
  \rlap{\textbf{Initial:}}\phantom{\textbf{Refinements:}}
  \insertDiagramWithBorder{outside}{diagrams/meta-op/start}{}

  \textbf{Refinements:}
  \vspace{-4ex}
  \begin{center}
    \insertDiagramWithBorderAnchor{north west}{outside-no-xsep}{diagrams/meta-op/provideValue1}{}
    \insertDiagramWithBorderAnchor{north west}{outside-no-xsep}{diagrams/meta-op/provideValue2}{}
    \insertDiagramWithBorderAnchor{north west}{outside-no-xsep}{diagrams/meta-op/provideValue3}{}

    \vspace{-2ex}
    \insertDiagramWithBorderAnchor{south west}{outside-no-xsep}{diagrams/meta-op/provideList1}{}
    \insertDiagramWithBorderAnchor{south west}{outside-no-xsep}{diagrams/meta-op/provideList3}{}

    {\Large\dots}
  \end{center}
  \caption{\label{fig:grammar} Example system of \metaName{}s (excerpt)}
\end{figure}

\section{Related work}

Many approaches for automatic program generation exist in the literature.
The field of inductive program synthesis tries to generate programs from concrete data in the form of argument-result pairs of the program that is to be synthesized.
Inductive synthesis techniques include, among others, inductive logic programming, evolutionary algorithms, and techniques based on term-rewriting and functional programming~\cite{kitzelmann2010}.

In machine learning program generation is treated as a natural-language processing task, i.e., given a verbal description of behavior produce source code of a program implementing that behavior~\cite{triet2020}.
Recent advances in large-language-models have improved code generation capabilities and allow for applications at larger scale.
This includes the generation of complete programming exercises using these models, although human supervision is still required to catch some inaccuracies~\cite{cambaz2024,sarsa2020}.

In contrast to the above-mentioned approaches our system does not actually create programs freely within the target language.
All concrete fragments are handwritten by the educator and the system only arranges these partial artifacts into a complete program.

Our approach is more similar to template-based systems for code generation.
For example, many approaches exist for generating program code from system descriptions in some modelling language~\cite{syriani2018}.
Templates, or template-like approaches, are also often used for automatic generation of exercise task variants~\cite{shah2002,thomas2019}.
Our system can be seen as a multi-level template approach, i.e., abstract implementations are templates where the holes (abstract idioms) get filled with other templates (concrete idioms).
In contrast to existing systems a central feature of our system is the generation of multiple related artifacts of different kinds.

Ask-Elle~\cite{gerdes2017}, a programming tutor system for Haskell, is capable of incrementally giving hints to students on how to proceed to solve a given programming task.
Different solution strategies are encoded in a strategy-DSL that defines possible refinement step~\cite{heeren2010}.
From these steps different sample solutions can be derived.

Our previous work on automatic generation of Haskell IO exercise tasks~\cite{wflp20-paper} requires random generation of both program code and specification expressions (in the language of \cite{tfpie19-paper}).
Previously we only used ad-hoc generation of randomized specifications and programs or attempted to derive programs from specifications.
The presented framework improves upon this by providing a principled approach to not only the generation of a single artifact but related artifacts that realize the same abstract intent.

From a formal perspective abstract idioms and abstract implementations can be viewed as objects and arrows in a multicategory~\cite{leinster2004}, also known as a colored operad.
Our substitution operation on diagrams then provides the composition operation on these arrows.
Viewed from this angle, the instantiation of abstract idioms with concrete ones becomes an algebra on this multicategory, i.e., a map into the multicategory of sets.
Multicategories and the closely related monoidal categories are frequently used to model a wide range of different systems of interconnected components and provide a graphical representation through corresponding wiring diagrams~\cite{foley2021,fong_spivak_2019,selinger2011,spivak2013}.

\section{Conclusion \& Future work}
We presented a framework for generating different related artifacts for use in automatic exercise generation.
By deriving all artifacts from a single source of abstract behavior description we can easily get a collection of connected artifacts without the need for complex analysis and lossy cross-translation of artifacts.
The central component of our system are abstract implementations to describe intended behavior.
Abstract implementations are given as wiring diagrams connecting multiple abstract idioms.
We introduced a simple mechanism to automatically transform these diagrams into variations that use specialized context dependent alternative implementation strategies.
By choosing concrete realizations for each abstract idiom we then create a complete artifact by composing the small parts according to the wiring structure of the abstract implementation.
Lastly, we showed how we can define a kind of context-free grammar producing abstract implementations.
This allows us to make the process of generating artifacts also choose between different behaviors instead of only different realizations of one specific behavior.

We have built a prototype implementation for the presented generation approach.
Currently, wiring diagrams need to be specified as one-dimensional expressions using sequential and parallel composition, but we plan building a graphical interface allowing for quicker and more intuitive creation of abstract implementations.

The ability of the presented system to generate diverse variations of programs and other artifacts requires
a sufficiently large collection of different concrete idioms for each abstract idiom used in the description of intent, as well as alternative implementations and merge-rules where applicable.
Creating such a collection requires more work than simply creating a handful of different artifacts manually.
In case we also generate intent itself with the help of the grammar-like abstract patterns the required effort is even higher, since we need to provide enough variability for every possible resulting intent.
But depending on context even a collection that enables generation of a few dozen different artifacts can be useful for automatic or manual exercise creation and variation.
Since one of our goals for the system is reusability, we can then gradually build up a larger collection over time.
Through the underlying combinatorics the payoff in added variations will also increase with growing collection size.

Replacing a program in an exercise template with another variant, either with identical or different behavior, can change the difficulty of the exercise tasks significantly.
Some programs are easier to read and comprehend than others, e.g., heavy use of higher-order functions can make programs shorter but might increase the difficulty of understanding them, especially for novice programmers.
Currently, we have no metrics or heuristics to anticipate the difficulty of an exercise task based on the programs (or other artifacts) it is built from.
However, automatically determining the difficulty of generated exercise tasks can be useful, especially in settings with no immediate oversight by an educator.

Since the presented framework is agnostic to the kinds of generated artifacts we also plan on experimenting with various different target artifacts.
Instead of complete programs one could generate programs with holes to be filled in by students or intentionally provide incorrect concrete idioms and require students to identify these mistakes.
But we could also leverage the framework to improve feedback for wrong student submissions.
Since we know the structure of the intended behavior, we could generate specifications, i.e., test cases, only for a ``prefix'' of the behavior, trying to determine where the given program's behavior diverges from a correct solution.
Lastly, artifact generation for other application domains besides Haskell IO, or maybe even beyond programming tasks, is possible.

\bibliographystyle{eptcs}
\bibliography{sources}

\end{document}